\documentclass[preprint,12pt]{elsarticle}

\usepackage{amsmath}
\usepackage{amssymb}
\usepackage{physics}
\usepackage{graphicx}
\usepackage{tikz}
\usepackage{array}
\usepackage{multirow}
\usepackage[frozencache=true,cachedir=.]{minted} 
\usepackage{hyperref}

\newcounter{bla}

\journal{Computer Physics Communications}

\begin{document}

\begin{frontmatter}



\title{TNQMetro: Tensor-network based package for efficient quantum metrology computations}


\author[a]{Krzysztof Chabuda}
\author[a]{Rafa{\l} Demkowicz-Dobrza{\'n}ski\corref{author}}

\cortext[author] {Corresponding author.\\\textit{E-mail address:} demko@fuw.edu.pl}
\address[a]{Faculty of Physics, University of Warsaw, ul. Pasteura 5, PL-02-093 Warszawa, Poland}

\begin{abstract}
TNQMetro is a numerical package written in Python for calculations of fundamental quantum bounds on measurement precision. Thanks to the usage of the tensor-network formalism it can beat the curse of dimensionality and provides an efficient framework to calculate bounds for finite size system as well as determine the asymptotic scaling of precision in systems where quantum enhancement amounts to a constant factor improvement over the Standard Quantum Limit. It is written in a user-friendly way so that the basic functions do not require any knowledge of tensor networks.
\end{abstract}


\end{frontmatter}



{\bf PROGRAM SUMMARY}

\begin{small}
\noindent
{\em Program Title:} TNQMetro \\
{\em Developer's repository link:} \url{https://github.com/kchabuda/TNQMetro} \\
{\em Licensing provisions:} MIT \\
{\em Programming language:} Python \\
{\em Nature of problem:}\\
Exponential growth of the Hilbert space dimension with the number of particles involved is a serious roadblock for numerical studies of the potential of quantum enhanced metrology. It leads to an exponential growth of the computational complexity of even most elementary quantum mechanical calculations, not to mention more advanced computational tasks, such as the ones required for studying the metrological potential of quantum states, e.g. computation of the quantum Fisher information (QFI). \\
{\em Solution method:}\\
Thanks to the use of the tensor-network formalism, where quantum states are represented as matrix product states and operators as matrix product operators, it is possible to obtain an efficient description where space complexity scales linearly with the number of elementary particles constituting the physical system. Furthermore, it is possible to efficiently optimize QFI over quantum states and operators in those representations, applying the ideas presented in [1]. This allows to study sophisticated quantum metrological models that are beyond the grasp of the standard numerical methods utilizing the full Hilbert space representation of quantum states and operations. \\

\end{small}

\section{Introduction} \label{sec:Introduction}
Tensor networks are well known in quantum many-body physics as efficient representation for problems with local structure of correlations~\cite{Orus2019}. In such problems they allow to bypass the curse of dimensionality (which arise from the exponential growth of the Hilbert space with the number of particles) and perform highly efficient calculations which e.g. for one-dimensional systems scales linearly with number of particles in the system~\cite{Schollwoeck2011}. Their properties have recently been employed in quantum metrology~\cite{Chabuda2020} to provide a complete framework for efficient calculations of the fundamental bounds on the precision of estimation of an unknown parameter of quantum dynamics in presence of locally correlated noise. The framework allows to determine these bounds not only for the finite many-body systems, but also in the asymptotic limit when the number of particles goes to infinity. The drawback is that the framework is not easy to implement, especially for the scientists without previous experience with the tensor networks. To mitigate this problem and allow a widespread use of the tensor-network based methods in quantum metrology community we have developed the TNQMetro (Tensor Networks for Quantum Metrology) package. It is a ready-to-use implementation of the framework described in the paper~\cite{Chabuda2020} in the form of a Python numerical package. Despite certain level of complexity of the package, the most basic and at the same time the most practically useful functions do not require from the end user any knowledge about tensor networks.

The structure of the paper is simple. After introduction in Sec.~\ref{sec:Introduction} we move to Sec.~\ref{sec:Background} where we give basic information about one-dimensional tensor networks which are the main object of TNQMetro and introduce the notion of the quantum Fisher information (QFI), optimization of which is the leitmotif of TNQMetro. Sec.\ref{sec:Main} is the main section where we introduce the basic functions of TNQMetro and explain how to encode a given quantum dynamics in the formalism of TNQMetro. In Sec.~\ref{sec:Example} we present the usage of TNQMetro to find bounds on precision in a phase estimation problem under different kinds of noise and provide insight into the performance of the algorithm. We finish with a short summary in Sec.~\ref{sec:Summary}.

\section{Background} \label{sec:Background}
The unique feature of the TNQMetro package is that it solves the problem of calculation and optimization of the QFI using the tensor-network representation involving matrix product states (MPS) and matrix product operators (MPO). They are one-dimensional tensor networks, i.e. chains, and can be used efficiently to represent vectors and operators in the Hilbert space provided their entanglement structure is simple enough~\cite{Eisert2010}. The description is in particularly efficient for states with short range entanglement structures. In the MPS representation, with open boundary conditions (OBC), a pure quantum state of $N$ distinguishable $d$-dimensional particles takes the form:
\begin{equation} \label{eq:MPS OBC}
	\ket{\psi} = \sum_{j_1, j_2 , \dots, j_N = 0}^{d-1} A\qty[1]^{j_1} A\qty[2]^{j_2} \dots A\qty[N]^{j_N} \ket{j_1 j_2 \dots j_N},
\end{equation}
so that the complex coefficient for each basis vector is obtained as a product of $N$ matrices $A\qty[n]^{j_n}$ of the dimension $D_{n-1} \cross D_{n}$, where at the start and at the end of the chain there is a covector and a vector, $D_0=D_N = 1$, so that the final result is indeed a scalar. $A[n]$ is a tensor of rank $3$ which for a given index $j_n \in \qty{0, \dots, d-1}$, called a physical index, is a matrix (tensor of rank $2$). The remaining two indices of $A[n]$ are called virtual indices and are labeled $\alpha_{n-1} \in \qty{0, \dots, D_{n-1}-1}$ and $\alpha_{n} \in \qty{0, \dots, D_{n}-1}$. The largest $D_{n}$ is called the bond dimension and we label it simply as $D$. It is an important parameter in the optimization procedure as it determine maximal entanglement-rank between the two parts of the studied system. Because, in case of OBC, $D_n$ for different $n$ are not the same, the above MPS~\eqref{eq:MPS OBC} is represented in TNQMetro as a list of length $N$ of numpy.ndarrays. If coefficients of the state~\eqref{eq:MPS OBC} are encoded in Python as a list \texttt{psi} then to call a specific element $\qty(A\qty[n]^{j})^{a}_{b}$ one should type \texttt{psi[n][a,b,j]}.

The definition of an MPS in the periodic boundary conditions (PBC) description reads:
\begin{equation} \label{eq:MPS PBC}
	\ket{\psi} = \sum_{j_1, j_2 , \dots, j_N = 0}^{d-1} \Tr(A\qty[1]^{j_1} A\qty[2]^{j_2} \dots A\qty[N]^{j_N}) \ket{j_1 j_2 \dots j_N},
\end{equation}
where the presence of the trace makes the first and the last index formally connected with each other. In typical scenarios, when working with PBC all $D_n$ will be equal. Therefore, when representing MPS with PBC in TNQMetro, we use a bit more efficient description---one bigger numpy.ndarray. So if coefficients of the state~\eqref{eq:MPS PBC} are encoded in Python as a numpy.ndarray \texttt{psi}, then to call specific element $\qty(A\qty[n]^{j})^{a}_{b}$ one should type \texttt{psi[a,b,j,n]}.

In most quantum metrology problems we are interested in behavior of a system in the limit of large number of particles when the boundary effects are negligible. Therefore it is not expected that there will be any significant difference in the physical results obtained when imposing OBC or PBC. However, the choice of a particular form of the boundary conditions plays an important role in tensor-network formalism. Contraction of states with OBC have much lower computational complexity (both in terms of time and space) and, moreover, for OBC it is possible to use a convenient canonical form~\cite{Orus2014}. TNQMetro supports both OBC and PBC, but if there is no physical reason to choose PBC in the problem considered it is recommended to use OBC for finite size systems (calculation for thermodynamic limit are by design independent of boundary conditions). In fact the basic TNQMetro functions are doing optimization using states and operators in OBC by default.

Using tensor networks it is possible to represent not only quantum states but also operators using the matrix product operators (MPO). A general description is very similar to MPS, and MPO differ from MPS only by an additional physical index. In particular, when written in the MPO OBC representation an operator $O$ takes the form:
\begin{equation} \label{eq:MPO OBC}
	O = \sum_{\substack{j_1, j_2 , \dots, j_N, \\ k_1, k_2 , \dots, k_N = 0}}^{d-1} A\qty[1]^{j_1}_{k_1} A\qty[2]^{j_2}_{k_2} \dots A\qty[N]^{j_N}_{k_N} \dyad{j_1 j_2 \dots j_N}{k_1 k_2 \dots k_N}.
\end{equation}
In Python we would refer to one of its specific entries $\qty(A\qty[n]^{j}_{k})^{a}_{b}$ as \texttt{O[n][a,b,j,k]} (assuming that all coefficients are saved in a list named \texttt{O}).

Now, we are going to present a simple example, showing how to transform a product state $\ket{\psi}$ into an MPS with OBC as well as PBC, following the TNQMetro convention for order of indices as presented above. When dealing with product states one only needs to reorder the indices in the standard state description in order to obtain an MPS representation with bond dimension $D=1$. Consider $N=1000$ two-level particles, where the state of a single particle is represented as a point on the equator of Bloch sphere (which is in a lot of cases a good ansatz for optimization for two-level problems):
\begin{equation} \label{eq:product}
	\ket{\psi} = \frac{1}{\sqrt{2^N}} \qty(\ket{0}+ \mathrm{e}^{\mathrm{i} \theta} \ket{1})^{\otimes N}.
\end{equation}
In this case, the MPS representation corresponds simply to $A[n]^0 = 1/\sqrt{2}$, $A[n]^1 = \mathrm{e}^{\mathrm{i} \theta}/\sqrt{2}$ (for each $n$) and the code in Python for this example is in the Listing~\ref{lst:mps example}.

\begin{listing}[h]
\begin{minted}[frame=lines]{python}
import numpy as np
N = 1000
theta = np.pi / 2
psi0 = np.array([1, np.exp(1j * theta)]) / np.sqrt(2)
psi_MPS_OBC = psi0[np.newaxis, np.newaxis, :]
psi_MPS_OBC = [psi_MPS_OBC] * N
psi_MPS_PBC = psi0[np.newaxis, np.newaxis, :, np.newaxis]
psi_MPS_PBC = np.tile(psi_MPS_PBC, (1,1,1,N))
\end{minted}
\caption{Matrix product state (MPS) representation of a state~\eqref{eq:product} for $N=1000$ and $\theta = \pi/2$ with open (OBC) and periodic (PBC) boundary conditions.}
\label{lst:mps example}
\end{listing}

A standard paradigm in quantum metrology~\cite{Giovannetti2006} is to think of some density matrix $\rho_0$, which describes initial state of the system, evolving through a quantum channel $\Lambda_\varphi$ (completely positive trace preserving map~\cite{Nielsen2000}) which is parameterized by an unknown parameter $\varphi$ (in this paper we are focusing on the problems with single unknown parameter). This way the information about the parameter is encoded in the density matrix at the output of the channel $\rho_\varphi = \Lambda_\varphi\qty[\rho_0]$. Now a measurement takes place and using the function called estimator $\hat{\varphi}\qty(\vb{x})$, the value of the unknown parameter is estimated based on the results of the measurement $\vb{x}$.

One of the main goals of quantum metrology and the TNQMetro package is to find a fundamental bound on the precision of estimation of an unknown parameter. In order to obtain a fundamental bound on the estimation variance $\Delta^2 \hat{\varphi}$ one would need to perform an optimization over all possible initial states, measurements and estimators. Fortunately, when following the frequentist approach to estimation, one may resort to the powerful quantum Cram{\'e}r-Rao bound which gives us lower bound on the estimation variance~\cite{Helstrom1976,Holevo1982}:
\begin{equation} \label{eq:QCR}
	\Delta^2 \hat{\varphi} \geq \frac{1}{F \qty[\rho_\varphi]},
\end{equation}
where $F \qty[\rho_\varphi]$ is the QFI. Usually QFI is defined as $F \qty[\rho_\varphi] = \Tr(\rho_\varphi L^2)$ where the Hermitian operator $L$ is called the Symmetric Logarithmic Derivative (SLD) and it is defined implicitly by the equation $\dot{\rho}_\varphi = \tfrac{1}{2} \qty(\rho_\varphi L + L \rho_\varphi)$ in which $\dot{\rho}_\varphi = \pdv*{\rho_\varphi}{\varphi}$. Here we are using an equivalent definition of the QFI~\cite{Macieszczak2013,Macieszczak2014}:
\begin{equation} \label{eq:QFI}
	F \qty[\rho_\varphi] = \max_{L} F \qty[\rho_\varphi, L], \quad F \qty[\rho_\varphi, L] = 2 \Tr(\dot{\rho}_\varphi L) - \Tr(\rho_\varphi L^2).
\end{equation}
which has the form of a quadratic optimization problem and as such this is easier to cast in the tensor-network formalism. Now, to obtain the fundamental bound we have to optimize the QFI over initial state $\ket{\psi_0}$ (it can be easily proven that optimal initial state is pure), so the final task takes the form of double optimization problem:
\begin{equation} \label{eq:QFI2}
	F = \max_{\ket{\psi_0}} F\qty[\Lambda_\varphi\qty[\ket{\psi_0}]] = \max_{\ket{\psi_0}} \max_{L} F\qty[\Lambda_\varphi\qty[\ket{\psi_0}], L],
\end{equation}
where $F\qty[\Lambda_\varphi\qty[\ket{\psi_0}], L]$, defined in Eq.~\eqref{eq:QFI}, is our figure of merit (FoM). It should be emphasized that the scope of applicability of the package described in this paper is not restricted to the frequentist approach but may also be used to obtain fundamental bounds within the Bayesian approach if the optimized quantity can be put in the form set by Eq.~(\ref{eq:QFI},~\ref{eq:QFI2}) (see e.g. the quantum Allan variance optimization problem~\cite{Chabuda2016}).

In realistic metrological scenarios, which take into account the effects of noise, the asymptotic scaling of the QFI will be linear in $N$, and the quantum enhancement will amount to a constant factor improvement \cite{Escher2011,DemkowiczDobrzanski2012} over the Standard Quantum Limit (SQL). As such, QFI will asymptotically be an extensive quantity. In TNQMetro this fact is used to introduce a procedure of renormalization and directly obtain asymptotic quantum enhancement coefficient for such systems.

We should note that TNQMetro has applications beyond the field quantum metrology, thanks to the fact that QFI is linked with fidelity of quantum states $\mathcal{F}$~\cite{Braunstein1994}:
\begin{equation} \label{eq:fid}
	\mathcal{F} \qty(\rho_\varphi, \rho_{\varphi + \varepsilon}) = 1 - \tfrac{1}{8} F \qty[\rho_\varphi] \varepsilon^2 + \order{\varepsilon^3},
\end{equation}
and, therefore, can be used in many-body physics studies where state fidelity is the quantity of interest, e.g. in studies of phase transitions.

\section{TNQMetro usage} \label{sec:Main}
TNQMetro is a package written in Python 3. It requires two external packages: NumPy and ncon\footnote{ncon is a Python 3 implementation of NCON function (written originally in MATLAB) and is used for tensor-network contraction~\cite{Pfeifer2014}.}. It can installed from the Python Package Index by typing the command \texttt{pip install tnqmetro} to the Python interpreter.

The main goal of the TNQMetro package is to find fundamental quantum bounds on precision by maximizing the expression~\eqref{eq:QFI2}. The optimization process is done iteratively and consists of several layers. The first layer involves the optimization of FoM over the operator $L$ or vector $\ket{\psi_0}$ (to differentiate between them we call optimization over $\ket{\psi_0}$ a dual problem and if we want to emphasize that we are focusing on the dual problem then we label figure of merit as FoMD). Each of those are expressed as an MPO/MPS and the optimization takes place on the level of each tensor in the chain separately, as described in details in the paper~\cite{Chabuda2020}. The second level is alternating the optimization of $L$ and $\ket{\psi_0}$ to realize the double optimization problem. The third level is to check the convergence of the FoM while the bond dimension of $L$ ($D_L$) and $\ket{\psi_0}$ ($D_{\psi_0}$) are increased---we start from $D_L = D_{\psi_0} = 1$ and incrementally increase them until FoM does not change more then some threshold (by default $1\%$) neither in $D_L$ nor in $D_{\psi_0}$. For the details of optimization we refer to the paper~\cite{Chabuda2020}. TNQMetro has a module structure so it allows advanced users to have access to each of those stages of optimization separately but basic functions outputs just the final result of all of those optimization layers combined.

\begin{table}
	\centering
	\begin{tabular}{m{1.75cm} | m{3.95cm} | m{2.8cm} | m{3.6cm} }
		& $\displaystyle \max_{\ket{\psi_0}} \max_{L} F\qty[\Lambda_\varphi\qty[\ket{\psi_0}], L]$ & $\displaystyle \max_{L} F\qty[\Lambda_\varphi\qty[\rho_0], L]$ & $\displaystyle \max_{L} F\qty[\rho_\varphi, L]$ \\
		\hline
		\multirow{2}{1.75cm}{general} & \texttt{fin\char`_gen()} & & \texttt{fin\char`_state\char`_gen()} \\
		& $\Lambda_{\varphi}$, $\dot{\Lambda}_{\varphi}$ or $\Lambda_{\varphi}$, $\Lambda_{\varphi + \varepsilon}$, $\varepsilon$ & & $\rho_{\varphi}$, $\dot{\rho}_{\varphi}$ or $\rho_{\varphi}$, $\rho_{\varphi + \varepsilon}$, $\varepsilon$ \\
		\hline
		\multirow{3}{1.75cm}{TI $\Lambda_{\varphi}$ with unitary $\varphi$} & \texttt{fin()} & \texttt{fin\char`_state()} & \\
		& $[\text{so}_1^\text{b},\text{so}_2^\text{b},\ldots]$, $h$, & $[\text{so}_1^\text{b},\text{so}_2^\text{b},\ldots]$, $h$, & \\
		& $[\text{so}_1^\text{a},\text{so}_2^\text{a},\ldots]$ & $[\text{so}_1^\text{a},\text{so}_2^\text{a},\ldots]$, $\rho_0$ & \\
	\end{tabular}
	\caption{Comparison of applications for the four main TNQMetro functions for finite (\texttt{fin}) approach, alongside their main inputs. We introduce $\text{so}_{i}^\text{b/a}$ as a label for local superoperator acting before/after unitary parameter encoding (generated locally by $h$). Each of those functions have a infinite variant (which gives asymptotic scaling of FoM)---they have similar names but with \texttt{inf} segment instead of \texttt{fin}.}
	\label{table:functions}
\end{table}

Our tensor-network methods have two approaches: finite (\texttt{fin}) and infinite (\texttt{inf}). In the finite approach number of sites in the chain of tensors is finite. Often we associate one site with one particle but the physical interpretation can differ from problem to problem. The infinite approach operates in the thermodynamic limit when the size of a system $N$ goes to infinity and we also assume that the whole system is translationally invariant (TI). However the main assumption is that for larger $N$ the FoM scales linearly with the size of a system and algorithm returns the asymptotic coefficient (which after multiplication by the size of a system gives the actual extensive FoM). In other words infinite approach assumes that because of noise the precision of a measurement follow the linear scaling with the number of elementary probes (SQL scaling). It is recommended to check first whether in the model studied, the FoM scales indeed linearly with the size of a system ($F = \text{const} * N$) on finite network before using the infinite approach---otherwise one may expect a diverging result in the infinite approach.

For each approach there are four main functions in TNQMetro package---see Tab.~\ref{table:functions}. For the brevity of the presentation we focus on the finite approach but each of those functions has an infinite variant with \texttt{inf} segment in the name instead of \texttt{fin}. The most general function which returns the fully optimized FoM is \texttt{fin\char`_gen()}. To specify a physical problem, a user has to provide information about the quantum channel $\Lambda_{\varphi}$ and its derivative (over estimated parameter $\varphi$) $\dot{\Lambda}_{\varphi}$ or a second channel $\Lambda_{\varphi + \varepsilon}$ which operates for the value of estimated parameter which is shifted by small parameter $\varepsilon$ (it is then used to compute a finite difference approximation to the actual derivative). The important remark is that the quantum channels have to be input in the form of superoperators\footnote{Superoperator $\Phi_\Lambda$ associated with a quantum channel $\Lambda$ is a operator which acts on a vectorized density matrix, so if $\rho = \Lambda\qty[\rho_0]$ then $\ket{\rho} = \Phi_\Lambda \ket{\rho_0}$.}. In some cases it is important to be able to compute a bound on precision for the specific state used in experiment. In such a scenario one can use the function \texttt{fin\char`_state\char`_gen()}. In this case we do not need information about the quantum channel itself but only about the density matrix at the end of the quantum channel $\rho_{\varphi}$, as well as its derivative $\dot{\rho}_{\varphi}$ or a second density matrix $\rho_{\varphi + \varepsilon}$ to create a discrete derivative approximation. Note that in both of the above described functions the superoperators or the density matrices have to be provided using the MPO representation.

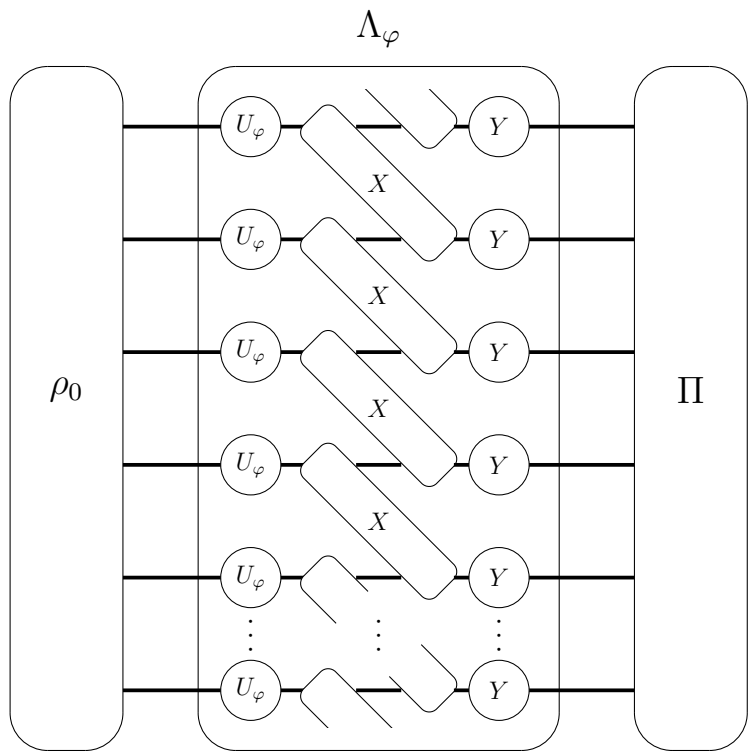
\begin{figure}
	\centering
	\begin{tikzpicture}
		\begin{scope}[rotate=-90]
		\draw (-1.3,1.7) node[scale=1.2] {$\Lambda_\varphi$};
		\draw (-0.3,-3.2) -- (7.8,-3.2);
		\draw (-0.3,-3.2) arc (270:180:0.5);
		\draw (7.8,-3.2) arc (-90:0:0.5);
		\draw (-0.3,-1.7) -- (7.8,-1.7);
		\draw (-0.3,-1.7) arc (90:180:0.5);
		\draw (7.8,-1.7) arc (90:0:0.5);
		\draw (-0.8,-2.7) -- (-0.8,-2.2);
		\draw (8.3,-2.7) -- (8.3,-2.2);
		\draw (3.5,-2.45) node[scale=1.2] {$\rho_0$};
		\draw (-0.3,-0.7) arc (270:180:0.5);
		\draw (-0.3,-0.7) -- (7.8,-0.7);
		\draw[line width=0.5mm] (0,-1.7) -- (0,0.8);	
		\foreach \x in {1.5, 3.0, ..., 7}
		{
			\draw[line width=0.5mm] (\x,-1.7) -- (\x,0.8);
		}
		\draw (6.8,0) node[rotate=-90] {$\dots$};
		\draw (7.8,-0.7) arc (-90:0:0.5);
		\draw[line width=0.5mm] (7.5,-1.7) -- (7.5,0.8);
		\foreach \x/\c in {0/{1}, 1.5/{2}, 3/{3}, 4.5/{4}, 6/{5}}
		{
			\draw[fill=white] (\x,0) circle (0.4);
			\draw (\x,0) node[scale=0.8] {$U_\varphi$};
		}
		\draw[fill=white] (7.5,0) circle (0.4);
		\draw (7.5,0) node[scale=0.8] {$U_\varphi$};
		\draw[rounded corners, fill=white, rotate around={45:(0.75,1.7)}] (-0.5,1.4) rectangle (2.0,2.0);
		\draw[rounded corners, fill=white, rotate around={45:(3.75,1.7)}] (2.5,1.4) rectangle (5.,2.0);
		\draw[rounded corners, fill=white, rotate around={45:(6.75,1.7)}] (5.5,1.4) rectangle (8.,2.0);
		\draw (0.75, 1.7) node[scale=0.8] {$X$};
		\draw (3.75, 1.7) node[scale=0.8] {$X$};
		\fill[white,rotate around={45:(6.75,1.7)}] (6.25,1.3) rectangle (7.25,2.1);
		\draw (6.8,1.7) node[rotate=-90] {$\dots$};
		\foreach \x in {0, 1.5, 3.0, ..., 8.5}
		{
			\draw[line width=0.5mm] (\x,1.4) -- (\x,2.0);
		}
		\draw[rounded corners, fill=white, rotate around={45:(2.25,1.7)}] (1.0,1.4) rectangle (3.5,2.0);
		\draw[rounded corners, fill=white, rotate around={45:(5.25,1.7)}] (4.0,1.4) rectangle (6.5, 2.0);
		\draw (2.25, 1.7) node[scale=0.8] {$X$};
		\draw (5.25, 1.7) node[scale=0.8] {$X$};
		\begin{scope}
			\clip (-0.5,0) rectangle (0.5,3);
			\draw[rounded corners, fill=white, rotate around={45:(-0.75,1.7)}] (-1.,1.4) rectangle (0.5,2.0);
		\end{scope}
		\begin{scope}
			\clip (7,0) rectangle (8,3);
			\draw[rounded corners, fill=white, rotate around={45:(8.25,1.7)}] (7,1.4) rectangle (8.5,2.0);
		\end{scope}
		\foreach \x in {0, 1.5, 3.0, ..., 8.5}
		{
			\draw[line width=0.5mm] (\x,2.7) -- (\x,3);
		}
		\foreach \x in {0, 1.5, 3.0, ..., 8.5}
		{
			\draw[line width=0.5mm] (\x,3.6) -- (\x,5.1);
		}
		\foreach \x in {0, 1.5, ..., 8.5}
		{
			\draw[fill=white] (\x,3.3) circle (0.4);
			\draw (\x,3.3) node[scale=0.8] {$Y$};
		}
		\draw (6.8,3.3) node[rotate=-90] {$\dots$};
		\draw (-0.3,4.1) arc (90:180:0.5);
		\draw (-0.3,4.1) -- (7.8,4.1);
		\draw (7.8,4.1) arc (90:0:0.5);
		\draw (-0.8,-0.2) -- (-0.8,3.6);
		\draw (8.3,-0.2) -- (8.3,3.6);

		\draw (-0.3,5.1) -- (7.8,5.1);
		\draw (-0.3,5.1) arc (270:180:0.5);
		\draw (7.8,5.1) arc (-90:0:0.5);
		\draw (-0.3,6.6) -- (7.8,6.6);
		\draw (-0.3,6.6) arc (90:180:0.5);
		\draw (7.8,6.6) arc (90:0:0.5);
		\draw (-0.8,5.6) -- (-0.8,6.1);
		\draw (8.3,5.6) -- (8.3,6.1);
		\draw (3.5,5.85) node[scale=1.2] {$\Pi$};
		\end{scope}
	\end{tikzpicture}
	\caption{A paradigmatic scheme of a quantum metrological problem with a quantum channel written in a tensor-network compatible formalism. Initial state $\rho_0$ evolves through the quantum channel $\Lambda_\varphi$, which is parameterized by an unknown parameter $\varphi$ and after the evolution the measurement $\Pi$ occurs. The whole channel is translationally invariant so it acts in the same way on each particle (each horizontal line represents a single particle). In our description, the quantum channel is constructed from multiple operation layers. In this example, the first layer represents the unitary encoding of the unknown parameter, $U_\varphi$, then we have two-particle operations $X$ and at the end one-particle operations $Y$ representing noise.}
	\label{fig:channel}
\end{figure}

In case of many important physical problems encountered in the field of quantum metrology (e.g. optical interferometry, Ramsey interferometry, magnetometry) an adequate description involves TI quantum channels with unitary parameter encoding. For such problems, there are two dedicated functions \texttt{fin()} and \texttt{fin\char`_state()} which give respectively the fully optimized FoM or its value for a specific input state. In these scenarios, a quantum channel is specified via layers of TI quantum operations---see Fig.~\ref{fig:channel} for an example. Because of the TI property, each layer is specified by one operation which is repeated within this layer---in the example from Fig.~\ref{fig:channel} these are: $U_\varphi$, $X$ (acts on two neighbouring particles) and $Y$. $U_\varphi$ is a special kind of operation---it is responsible for the unitary encoding of information about the estimated parameter $\varphi$ onto the state. In the current version of TNQMetro, there can be only one layer of such operations and we assume that $U_\varphi = \exp(- \mathrm{i} h \varphi)$, so the user has to specify only the local generator $h$ ("Hamiltonian"). For the other layers user needs to specify one defining operator in a form of a local superoperator\footnote{If one of these operations is described by a set of local Kraus operators then the function \texttt{Kraus\char`_to\char`_superoperator()} can be used to convert this set of local Kraus operators into a local superoperator.}. The final requirement for the user is to specify the order of those layers in the form of list of operations which act before the unitary encoding, then specify the generator $h$ and finally specify the list of operations that act after the unitary encoding---for the example from Fig.~\ref{fig:channel} this would be: \texttt{[], h, [X,Y]}. In case of the \texttt{fin\char`_state()} function one needs to additionally specify the initial state $\rho_0$ as an MPO. It is worth to mention that these two functions use auxiliary functions\footnote{\texttt{fin\char`_create\char`_channel()}, \texttt{fin\char`_create\char`_channel\char`_derivative()} and \texttt{channel\char`_acting\char`_on\char`_operator()}.} to prepare superoperators and density matrices in the MPO representation for the whole system from those small building block and then call the more general functions \texttt{fin\char`_gen()} and \texttt{fin\char`_state\char`_gen()}, already discuss before. As such, they may be regarded as a kind of interface functions that help the user provide the input for the FoM optimization procedures in a more convenient way.

To summarize the description, the \texttt{fin()} and \texttt{inf()} are the key functions of the TNQMetro package from the point of view of the end user. They are suitable to deal with the most relevant physical problems and do not require any knowledge about tensor networks. They allow to calculate the fundamental bounds for the precision of estimation for finite systems and the asymptotic scaling for systems following SQL. They have the same main inputs and require from the user to specify only local superoperators and the local generator of parameter encoding.

We would like to also mention the function \texttt{fullHilb()}. It has the same purpose and inputs as \texttt{fin()} but optimize the QFI using standard diagonalization in the full Hilbert space instead of utilizing tensor-network formalism. It can be used to benchmark the results obtained via tensor-network procedures but only for very small systems---not only because of the need of diagonalization of large density matrix representing the quantum state but also because the description of the quantum channel via the corresponding superoperator requires square of the amount of memory that a density matrix takes.

\section{Example: phase estimation} \label{sec:Example}
In this section we are going to present a practical application of TNQMetro to the problem of quantum phase estimation. It is a paradigmatic problem capturing the essence of all interferometric-like experiments~\cite{DemkowiczDobrzanski2015}. One can think of e.g. $N$ photons in the Mach–Zehnder interferometer, $N$ two-level atoms in the Ramsey interferometry experiment or multiple spin-$\frac{1}{2}$ particles sensing magnetic field. Each particle is a two level system, with states $\ket{0}$ and $\ket{1}$ on which a relative phase difference $\varphi$ is being imprinted. Hence, if a particle starts in the state $\ket{\psi_0} = \frac{1}{\sqrt{2}} \qty(\ket{0}+\ket{1})$, and there is no noise, then at the end of experiment the state of the particle becomes $\ket{\psi_\varphi} = \frac{1}{\sqrt{2}} \qty(\ket{0}+\mathrm{e}^{- \mathrm{i} \varphi} \ket{1})$ (up to an irrelevant global phase factor). This is equivalent to saying that a unitary operator $U_\varphi = \exp(- \mathrm{i} h \varphi)$, where the Hermitian generator ("Hamiltonian") is $h = \dyad{1}$, acts on each particle. Given $N$ particles, our goal will be to find the fundamental bound on the precision of estimation of $\varphi$, by optimizing the QFI over all possible initial states.

\subsection{Phase estimation without noise}
The simplest case is when there is no noise. Then, the corresponding quantum channel $\Lambda_\varphi$ can be described as:
\begin{equation}
	\rho_\varphi = \Lambda_\varphi[\rho_0] = \mathrm{e}^{ - \mathrm{i} H \varphi }\rho_0 \mathrm{e}^{\mathrm{i} H \varphi}, \quad H = \sum_{n=1}^N h^{[n]},
\end{equation}
where $h^{[n]}$ is the generator acting on the $n$th particle---in this case the quantum channel on Fig.~\ref{fig:channel} would consist only of $U_\varphi$ operations. QFI for this scenario can be calculated analytically, and the optimal initial state for the whole system is the NOON/GHZ state~\cite{Bollinger1996,Greenberger1989} for which $F = N^2$. But even for such simple case direct numerical optimization of the QFI using standard methods (diagonalization in full Hilbert space) is possible only for up to $~20$ qubits on a standard PC. Using TNQMetro we are capable of optimize QFI for $>1000$ qubits. To optimize the QFI for a particular $N$ (e.g. $N=1000$) we use the \texttt{fin()} function as presented in the Listing~\ref{lst:without noise}. This function outputs: the optimal value of QFI (\texttt{F}), matrix of values of QFI as a function of bond dimensions (\texttt{F\char`_m}), the optimal SLD in MPO representation (\texttt{L\char`_MPO}) and the optimal state in the MPS representation (\texttt{psi\char`_MPS}). For $N=1000$ the \texttt{fin()} returns $\mathtt{F}=994540$ which is in perfect agreement with the analytical result $F = N^2$ taking into account that by default the relative precision of this optimization is set to be around $1\%$.
\begin{listing}
\begin{minted}[frame=lines]{python}
import numpy as np
import tnqmetro
N = 1000 # number of sites in tensor-network
d = 2 # dimension of local Hilbert space
h = np.arange(d)
h = np.diag(h) # local generator ("Hamiltonian")
F, F_m, L_MPO, psi_MPS = tnqmetro.fin(N, [], h, [])
\end{minted}
\caption{Optimization of QFI using TNQMetro for $N=1000$ qubits with OBC for the problem of phase estimation without noise.}
\label{lst:without noise}
\end{listing}

\subsection{Phase estimation with uncorrelated noise}
The next case, on which we are going to focus, is phase estimation in presence of uncorrelated dephasing noise. The effect of dephasing noise is the loss of coherence between parts of superposition. On the level of density matrices it causes the diminishing of the off-diagonal elements. For such a meteorological problem the quantum channel $\Lambda_\varphi$ can be described as:
\begin{equation}
	\rho_\varphi = \Lambda_\varphi[\rho_0] = \mathrm{e}^{ - \mathrm{i} H \varphi } \Lambda[\rho_0] \mathrm{e}^{\mathrm{i} H \varphi}, \quad H = \sum_{n=1}^N h^{[n]},
\end{equation}
where $\Lambda$ denotes the part of the channel responsible for the effects of decoherence. In the case of dephasing noise it has the form:
\begin{equation} \label{eq:uncorr noise channel}
	\Lambda\qty[\rho_0] = \sum_{\mathbf{j}, \mathbf{k}} \mel{\mathbf{j}}{\rho_0}{\mathbf{k}} \mathrm{e}^{-\frac{c_1}{2} \norm{\mathbf{j} - \mathbf{k}}^2} \dyad{\mathbf{j}}{\mathbf{k}},
\end{equation}
where $\ket{\mathbf{j}} = \ket{j_1, j_2, \dots, j_N}$ ($j_i\in \{0,1\}$), is a basis states ($h^{\qty[n]} \ket{\mathbf{j}} = j_n \ket{\mathbf{j}}$), $c_1$ is a parameter describing the strength of the uncorrelated noise, whereas $\| \cdot \|$ is the standard vector norm for the vectors representing a given basis states i.e. 
$\mathbf{j} = \qty(j_1, j_2, \dots, j_N)^\mathrm{T}$ for $\ket{\mathbf{j}}$.

In this case the quantum channel in Fig.~\ref{fig:channel} would consist of unitary $U_\varphi$ operations and a single-particle operations $Y$ representing uncorrelated dephasing---see the paper~\cite{Chabuda2020} for the derivation of the $Y$ superoperator from the Eq.~\eqref{eq:uncorr noise channel}. For this problem there is no known analytical solution, but there exist methods to obtain bounds on the QFI (tight in the asymptotic regime)~\cite{Escher2011,DemkowiczDobrzanski2012} which give $F/N=\mathrm{e}^{-c_1} / \qty(1-\mathrm{e}^{-c_1})$ for the optimal state. For the problems with dephasing noise, the QFI follows the SQL scaling (it is linear in $N$) so apart from the computation for finite particle numbers we can also use the \texttt{inf()} function to directly calculate the asymptotic coefficient of the QFI---see Listing~\ref{lst:uncorrelated noise} for the code to calculate QFI for $N=1000$ as well as the asymptotic scaling for the phase estimation in the presence of uncorrelated noise of strength parameter $c_1 = 1$. Comparing the calculated asymptotic coefficient \texttt{F\char`_i}$=0.574$ with the exact value $F=\mathrm{e}^{-1} / \qty(1-\mathrm{e}^{-1}) \approx 0.582$ we see that it is about $1.4\%$ bellow which remains within the expected tolerance regime taking into account the $~1\%$ relative precision imposed during the optimization process. Result from the finite approach is about $5\%$ bellow the exact asymptotic value indicating that in order to reach the asymptotic limit, $N$ has to be even larger than $1000$.
\begin{listing}
\begin{minted}[frame=lines]{python}
import numpy as np
import scipy.linalg
import tnqmetro
N = 1000 # number of sites in tensor-network
d = 2 # dimension of local Hilbert space
h = np.arange(d)
h = np.diag(h) # local generator ("Hamiltonian")
c1 = 1. # uncorrelated noise strength parameter
aux = np.kron(h, np.eye(d)) - np.kron(np.eye(d), h)
# Y - local superoperator for uncorrelated noise
Y = scipy.linalg.expm(-c1 * aux @ aux / 2)
F_f, F_m_f, L_MPO_f, psi_MPS_f = tnqmetro.fin(N, [], h, [Y])
F_i, F_m_i, L_MPO_i, psi_MPS_i = tnqmetro.inf([], h, [Y])
\end{minted}
\caption{Optimization of QFI using TNQMetro for $N=1000$ qubits with OBC and in the asymptotic regime for the problem of phase estimation with uncorrelated noise.}
\label{lst:uncorrelated noise}
\end{listing}

\subsection{Phase estimation with correlated noise}
Now, we take into account the possibility of correlation in the noise. We are going to focus on the case when there are correlations between noise acting on the particles that ate the nearest neighbours. In order to take the correlations into account we have to use a more general description of the dephasing channel:
\begin{equation} \label{eq:corr noise channel}
	\Lambda\qty[\rho_0] = \sum_{\mathbf{j}, \mathbf{k}} \mel{\mathbf{j}}{\rho_0}{\mathbf{k}} \mathrm{e}^{-\frac{1}{2} \qty(\mathbf{j} - \mathbf{k})^\mathrm{T} C \qty(\mathbf{j} - \mathbf{k})} \dyad{\mathbf{j}}{\mathbf{k}},
\end{equation}
where $C$ is a noise correlation matrix:
\begin{equation}
	C = \begin{pmatrix}
		c_1    & c_2 & 0   & \dots  & 0    \\
		c_2    & c_1 & c_2 &        &        \\
		0      & c_2 & c_1 &        &        \\
		\vdots &     &     & \ddots & \vdots \\
		0    &     &     & \dots  & c_1    \\
	\end{pmatrix},
\end{equation}
where $c_1$ and $c_2$ are the parameters describing respectively the strength of uncorrelated and correlated parts of the noise. See that in the top-right and bottom-left corner of the matrix there are zeros---this is corresponds to the OBC case, while for PBC we would put there $c_2$. Of course, we could add even longer range correlations in the noise by adding $c_3$ on the next super- and subdiagonal but this makes bond dimension considerably bigger and increases the complexity of optimization. This case is depicted in Fig.~\ref{fig:channel} where $Y$ is the single-particle operations describing uncorrelated part of the noise and $X$ is the two-particle operation describing the effect of correlated part of the noise---the full derivation of the $Y$ and $X$ superoperators is presented in \cite{Chabuda2020}. Tensor based approach is right now the only method which can properly take into account the effects of correlations in noise and optimize the QFI for large systems. In \cite{Chabuda2020} the full analysis of the effects of correlations on the QFI in the phase estimation is given in the context of magnetic field sensing---Fig. 2a from~\cite{Chabuda2020} depicts how finite regime results approach the asymptotic value with increasing $N$, when calculated using a prototype version of the TNQMetro (written in MATLAB) and how the other state-of-the-art methods fail to provide satisfactory results in this case. The exact code allowing to calculate QFI for $N=1000$ and the asymptotic coefficient for this case with noise parameters $c_1=1$ and $c_2 = 0.1$ is provided in the Listing~\ref{lst:correlated noise}.
\begin{listing}[t]
\begin{minted}[frame=lines]{python}
import numpy as np
import scipy.linalg
import tnqmetro
N = 1000 # number of sites in tensor-network
d = 2 # dimension of local Hilbert space
h = np.arange(d)
h = np.diag(h) # local generator ("Hamiltonian")
c1 = 1. # uncorrelated noise strength parameter
c2 = 0.1 # correlated noise strength parameter
aux = np.kron(h, np.eye(d)) - np.kron(np.eye(d), h)
# Y - local superoperator for uncorrelated noise
Y = scipy.linalg.expm(-c1 * aux @ aux / 2)
# X - local superoperator for correlated noise
X = np.kron(aux, aux)
X = scipy.linalg.expm(-c2 * X)
F_f, F_m_f, L_MPO_f, psi_MPS_f = tnqmetro.fin(N, [], h, [X,Y])
F_i, F_m_i, L_MPO_i, psi_MPS_i = tnqmetro.inf([], h, [X,Y])
\end{minted}
\caption{Optimization of QFI using TNQMetro for $N=1000$ qubits with OBC and in the asymptotic regime for the problem of phase estimation with correlated noise.}
\label{lst:correlated noise}
\end{listing}

\subsection{Performance}
\begin{figure}
	\centering
	\includegraphics[width=\textwidth]{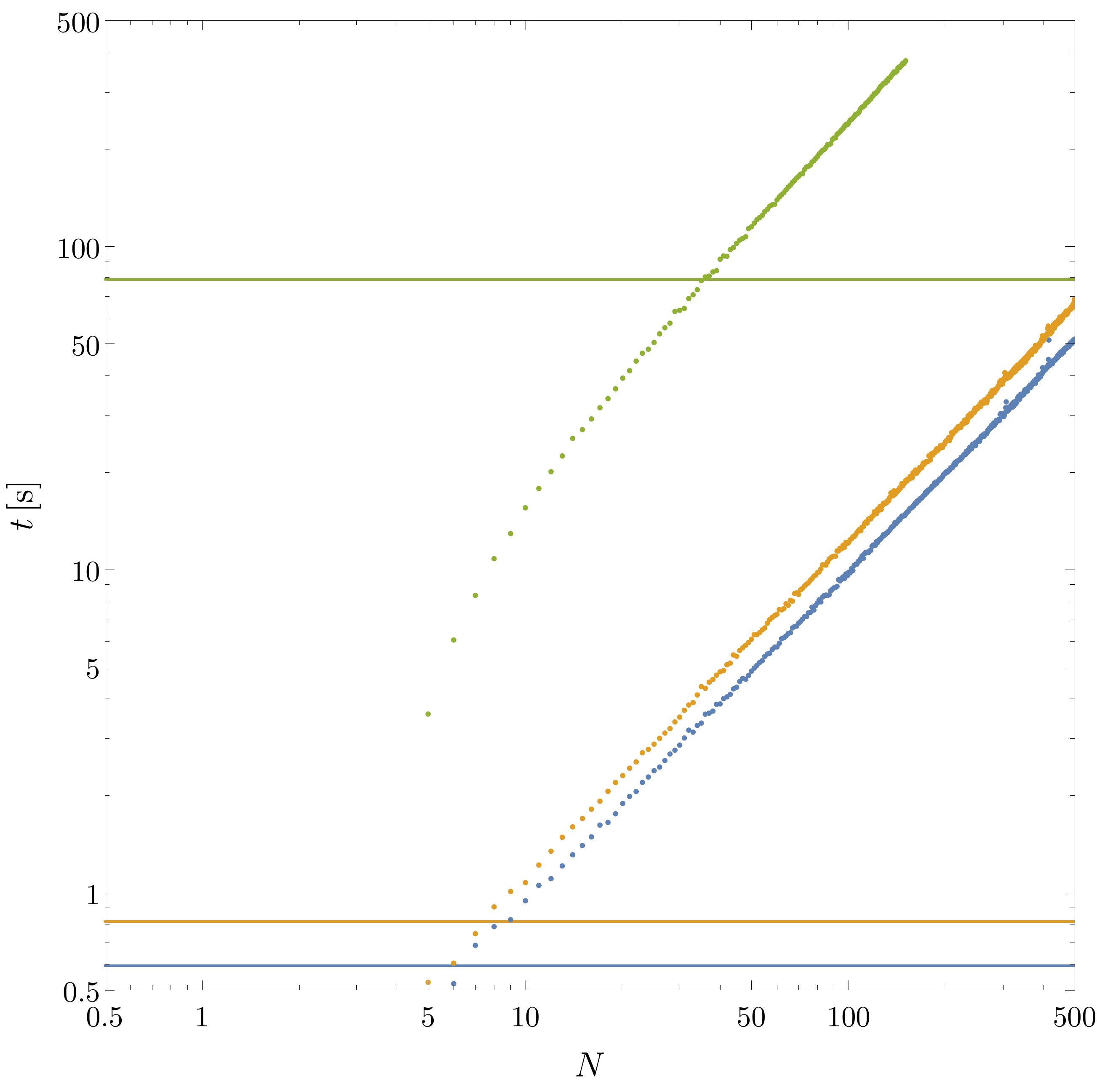}
	\caption{Time of QFI optimization $t$ for phase estimation with dephasing noise as a function of the chain length $N$. Results for three cases: uncorrelated noise (blue), noise which correlates two nearest particles (orange) and noise which correlates three nearest particles (green). Dots represent results for the finite chain length approach. Straight horizontal lines show how much time the infinite approach takes to optimize QFI with comparable precision.}
	\label{fig:perf}
\end{figure}

Bond dimension and the length of the chain (for finite approach) are the main factors impacting the time and memory complexity of the algorithm. Thanks to the use of tensor-network based methods the complexity scales roughly linearly with length of the chain $N$ (in contrast to exponential scaling of standard full Hilbert space methods). In Fig.~\ref{fig:perf} we show how much time the optimization takes as a function of the chain length $N$, for the phase estimation example with dephasing noise ($c_1=1$, $c_2 = 0.1$ and $c_3 = 0.01$). We present results when noise is: uncorrelated, correlates two nearest particles or correlates three nearest particles. To ensure that the observed differences results from the optimization process and are not affected by potentially increasing bond dimensions, we set that all calculations (for finite approach) to end up with $D_L = 2$ and $D_{\psi_0} = 3$ (which in this case resulted in relative imprecision of QFI to be under $4\%$). When we fit straight lines to the data (in log-log scale) the slopes are: $1.05$ (uncorrelated noise), $1.04$ (noise which correlates two nearest particles) and $1.19$ (noise which correlates three nearest particles). The first two cases are very close to perfectly linear scaling (slope equal to $1$). The last one shows some departure which we attribute to the fact that in this case matrices are large enough that garbage collection has to occur frequently and has substantial impact on the time of optimization.

The second most important factor impacting performance of the algorithm is the bond dimension $D$. Each MPS/MPO has its own bond dimension---the bigger the more correlated state/operator it can describe. Bond dimensions have significant impact on memory complexity of the algorithm. The most significant contribution to the memory complexity comes from the two tensors which we have to create at the beginning of each round of optimizations of the chain of tensors (involving operator $L$ or vector $\ket{\psi_0}$). Those two tensors combined have $D_{\psi_0}^4 \qty(D_\Lambda^2 D_L^4 + D_{\dot{\Lambda}}^2 D_L^2) \qty(N-1)$ elements\footnote{The factor $\qty(N-1)$ could be omitted but this would significantly increase time complexity of the algorithm.} (for OBC it can be a bit less because size of tensors can vary between sites). $D_\Lambda$ and $D_{\dot{\Lambda}}$ are bond dimensions for the MPO representing respectively quantum channel and its derivative (both as superoperators). For the case of TI channels with unitary parameter encoding (last row in Tab.~\ref{table:functions}): $D_{\dot{\Lambda}} = 2 D_{\Lambda}$ and $D_{\Lambda} = \prod_s D^{(s)}$ where $D^{(s)}$ is the number of non-zero singular values that will appear when considering the $s$-particle local operation---the upper bound on $D^{(s)}$ is $d^{2(s-1)}$ but e.g. for dephasing noise $D^{(s)} = \qty(2d-1)^{s-1}$. Time complexity is determined mainly by tensor contractions but because of complex network of contractions of different tensors the detailed analysis of the impact of bond dimensions on the time complexity is beyond the scope of this paper.

\section{Summary} \label{sec:Summary}
TNQMetro is easy to use and versatile numerical package which can be used in a variety of metrological problems to calculate fundamental bound on the precision of estimation. Thank to the usage of tensor networks it allows to study complex quantum channels and large many-body quantum systems.

\section*{Declaration of competing interest}
The authors declare that they have no known competing financial interests or personal relationships that could have appeared to influence the work reported in this paper.

\section*{Acknowledgements}
This work was supported by the National Science Center (Poland) grant No. 2016/22/E/ST2/00559.





\bibliographystyle{elsarticle-num}
\bibliography{bibliography}







\end{document}